\documentclass[%
 reprint,
superscriptaddress,
 amsmath,amssymb,
 aps,
prb,
]{revtex4-2}
\usepackage{ulem}
\usepackage{xcolor}
\usepackage{graphicx}
\usepackage{dcolumn}
\usepackage{bm}
\usepackage{gensymb}

\begin{document}

\preprint{prb}

\title{Wavevector analysis of plasmon-assisted distributed nonlinear photoluminescence along Au nanowire antennas}

\author{Deepak K. Sharma}
\email{deepak-kumar.sharma@u-bourgogne.fr}
\affiliation{Laboratoire Interdisciplinaire Carnot de Bourgogne, UMR 6303 CNRS, Université de Bourgogne Franche-Comté, 9 Avenue Alain Savary, 21000 Dijon, France}
\affiliation{Department of Physics, Indian Institute of Science Education and Research (IISER) Pune - 411008, India}
\author{Adrian Agreda}
\author{Julien Barthes}
\author{Gérard Colas des Francs}
\affiliation{Laboratoire Interdisciplinaire Carnot de Bourgogne, UMR 6303 CNRS, Université de Bourgogne Franche-Comté, 9 Avenue Alain Savary, 21000 Dijon, France}
\author{G. V. Pavan Kumar}
\affiliation{Department of Physics, Indian Institute of Science Education and Research (IISER) Pune - 411008, India}
\author{Alexandre Bouhelier}
\email{alexandre.bouhelier@u-bourgogne.fr}
\affiliation{Laboratoire Interdisciplinaire Carnot de Bourgogne, UMR 6303 CNRS, Université de Bourgogne Franche-Comté, 9 Avenue Alain Savary, 21000 Dijon, France}

\begin{abstract}
We report a quantitative analysis of the wavevector diagram emitted by nonlinear photoluminescence generated by a tightly focused pulsed laser beam and distributed along Au nanowire via the mediation of surface plasmon polaritions. The nonlinear photoluminescence is locally excited at key locations along the nanowire in order to understand the different contributions constituting the emission pattern measured in a conjugate Fourier plane of the microscope. Polarization-resolved measurements reveal that the nanowire preferentially emits nonlinear photoluminescence polarized transverse to the long axis at close to the detection limit wavevectors with a small azimuthal spread in comparison to the signal polarized along the long axis. We utilize finite element method to simulate the observed directional scattering by using localized incoherent sources placed on the nanowire. Simulation results faithfully mimic the directional emission of the nonlinear signal emitted by the different portions of the nanowire.

\end{abstract}

\maketitle


\section{\label{sec:level1}Introduction}

Metal nanostructures confine electromagnetic fields in the subwavelength regime by controlling resonant modes known as surface plasmons polaritons (SPP)~\cite{ref1,ref2}. The very large electric field present at the metal surface makes plasmonic-based devices excellent candidates to harness weak light-matter interactions~\cite{ref3,ref28} and to enhance optical nonlinear signals at the nanoscale~\cite{ref4,ref5}. Hand in hand to the electric field enhancement, a plasmonic nanostructure can direct the light according to the structural modes it supports~\cite{ref9,ref10,ref11,novotny}. For instance, placement of multiple elements in a specific spatial arrangement can provide unidirectional emission according to phase retardation produced by the whole geometry. A combination of plasmonic nanostructures placed in the form of Yagi-Uda design \cite{ref12,ref13} and metasurfaces \cite{ref14} have been utilized to provide high directivity and beaming effect from dipolar emitters.\\
In this context, a metal nanowire is an interesting plasmonic nanostructure that can confine, guide, and route SPP~\cite{ref15,ref16,ref17}. Plamonic nanowires confine surface plasmons well below the diffraction limit in the transverse dimensions and propagate the mode up to several micrometer long distance along its main axis. The SPP may then outcouple to free-space photons emitted in a defined angular direction when it scatters from the extremity of the metal waveguide~\cite{ref15}. For instance, plasmonic nanowires have been utilized for collecting and directing fluorescence signals emitted by nanocrystals~\cite{ref18} and quantum emitters~\cite{ref19} mainly in a direction imposed by the along the main axis. \\
Another interesting property brought by the large enhancement of the electric field associated with the excitation of the surface plasmon is the ability of the metal structure to generate its own surface nonlinearities~\cite{ref21,ref22}. We have recently showed that Au nanowires (AuNWs), and more generally extended 2D structures, may produce a distributed nonlinear photoluminescence (N-PL) when excited locally by a tightly focused ultrafast laser beam~\cite{ref23, kumar2018}. This nonlinear signal holds potential for developing advanced functionalities such as wavelength conversion~\cite{hasan} and all-optical Boolean operations~\cite{kumar2018, viarbitskaya2013}. Understanding this nonlinear response spatially transported in the geometry brings strategies on how to control it. Equally important is the knowledge about the direction the frequency-converted photons take when they scatter out of the structure and travel in free space.\\
With this motivation, we perform an analysis of the in-plane wavevector distribution of the N-PL developing on AuNW. We locally excite the N-PL at the extremity and at the center of a AuNW in order to understand and discriminate contributions from the different sections of the nanowire acting as an optical antenna.  We analyze the resulting Fourier plane pattern emitted by the distributed N-PL as a function of its polarization. We further use finite element method based simulations to understand our experimental results.

\section{\label{Sec:level2} Results and discussion}

\subsection{Experimental procedures}
\subsubsection{Fabrication of the Au nanowires}
\begin{figure}[b]
\includegraphics{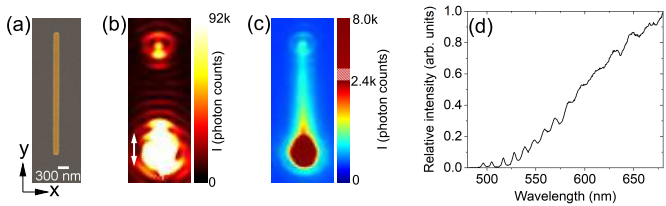}
\caption{\label{fig:1} (a) A scanning electron micrograph of a 3.5 $\mu$m long AuNW. The AuNW is aligned along the $y$-axis. (b) A  wide-field real plane image, showing surface plasmon propagation launched using a pulsed laser at $\lambda_0=808$ nm focused at the nanowire's bottom end (saturated region). The white arrow indicates the polarization of the incident laser beam along the axis of AuNW. (c) A false-color wide-field real plane image showing spectrally filtered delocalized N-PL from the AuNW. (d) The spectrum of N-PL recorded at the excitation location. The oscillations at shorter wavelengths are due to the transmission of the filter used.}
\end{figure}

We have fabricated 3.5 $\mu$m long and 160 nm wide polycrystalline AuNW on a glass substrate using a top-down fabrication procedure. Fabrication steps involve standard electron beam lithography (Pioneer, Raith GmbH) followed by metal depositions and a lift-off process. The subsequent metal layers consist of a 3 nm thick adhesive layer of titanium deposited by electron-beam physical vapor deposition (MEB 400, Plassys) and a 50 nm thick Au layer deposited by thermal evaporation. Figure \ref{fig:1}(a) presents a scanning electron micrograph of the AuNW.

\subsubsection{Optical measurements}
The AuNW and the glass coverslip are then placed on an inverted optical microscope (Nikon, Eclipse). The nanowire is optically excited at a wavelength of $\lambda_0$ = 808 nm at one of its extremities by a femtosecond pulsed laser beam (Coherent, Chameleon). The beam is tightly focused through the glass by a high numerical aperture (NA) oil-immersion objective lens (Nikon, $\times$100, NA = 1.49). The same objective lens is utilized for the collection of light scattered and emitted by the nanowire. Using relay lenses placed after the exit port of the microscope, the collected light is sent to a charge-coupled device (CCD) camera (Andor, Luca) or a spectrometer (Andor, Shamrock 303i). We image on the CCD the back focal plane of the objective lens using relay optics in a $4-f$ configuration. This maps the projected in-plane wavevector distribution of the light scattered or emitted by the nanowire in a conjugate Fourier plane~\cite{ref24}. An extra lens is inserted to capture corresponding wide-field real plane images. A piezo-electric stage (MadCity Labs, Nano LP100) is utilized for the precise displacement of the nanowire within the laser beam focus. Tight focusing and geometrical discontinuity at the end of the nanowire provide enough momentum for an efficient generation of SPP in the AuNW for an incident linear polarization along the long-axis of the nanowire ($y$-axis). 

SPP propagation at $\lambda_0$ is exemplified in the wide-field real plane image of Fig. \ref{fig:1}(b). The image is here taken at the laser wavelength and the excitation appears as a saturated intensity region at the bottom end of the nanowire. The weaker luminous response at the distal end results from SPP scattering at the physical discontinuity and confirms excitation and propagation of the SPP at the pump wavelength. 

Next, we increase the intensity of the laser to $\sim$ 7.5 GW/m$^2$ for triggering a SPP-mediated nonlinear response of the nanowire. In particular, we are spectrally selecting the nonlinear photoluminescence generated and developing in the AuNW and rejecting the second harmonic generation. N-PL is spectrally filtered from the laser beam using a dichroic beam splitter (Chroma) and a set of filters (Thorlabs) placed in the collection path. A spectrally filtered wide-field real plane image in Fig. \ref{fig:1}(c) shows the distributed N-PL over the whole structure~\cite{ref23,ref21}, with the strongest response at the excitation area. N-PL is an incoherent emission process~\cite{boyd1986} and consists of a wavelength continuum as shown by the raw spectrum taken at excitation end of the AuNW in Fig \ref{fig:1}(d). The spectrum in Fig. \ref{fig:1}(d) is limited by the mentioned spectral filters set. 

\subsection{N-PL generation in the AuNW antenna and its quantitative wavevector analysis}

\begin{figure}
\includegraphics{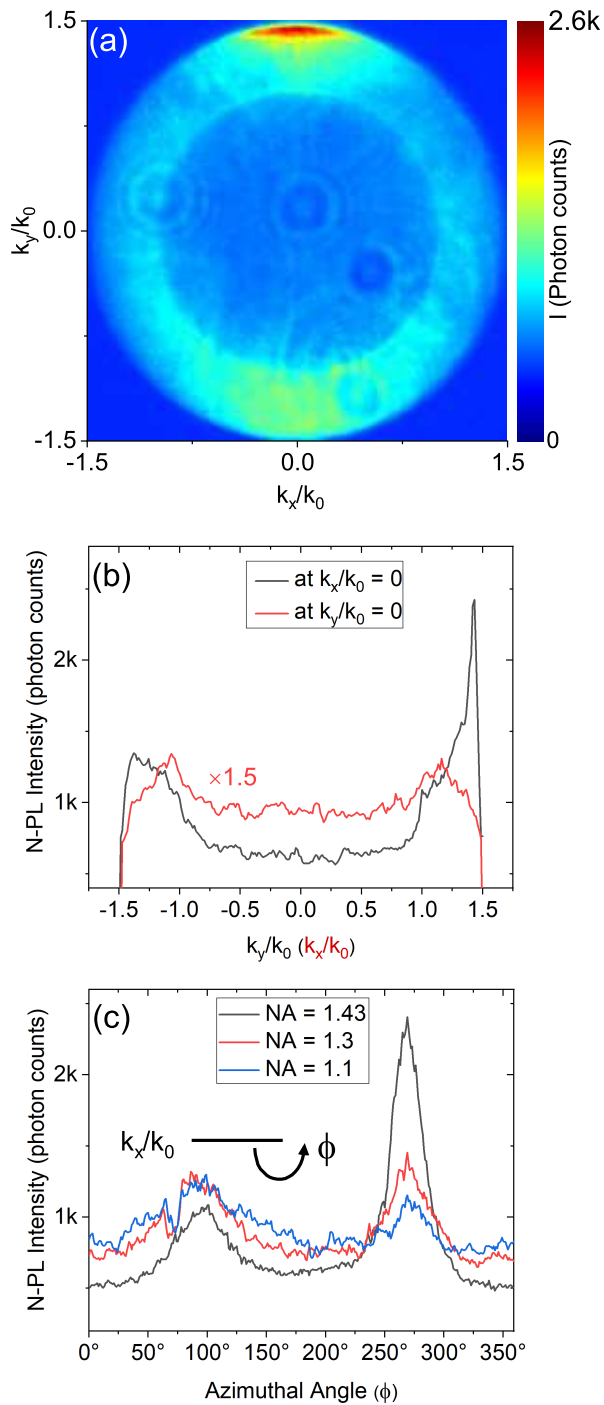}
\caption{\label{fig:2} (a) Fourier plane image of the delocalized N-PL corresponding to the real plane image shown in Fig. \ref{fig:1}(c). (b) N-PL intensity plots along $k_y$/$k_0$ (black) and $k_x$/$k_0$ (red) axes of the Fourier plane image. (c) N-PL intensity distribution profile along the azimuthal coordinate ($\phi$) plotted at radial coordinates 1.1, 1.3, and 1.43.}
\end{figure}

Next, we project the back focal plane of the objective lens onto the imaging camera to understand the wavevector distribution of the distributed N-PL~\cite{ref24}. Figure \ref{fig:2}(a) shows the Fourier plane pattern of the distributed N-PL. Axes are in unit of numerical aperture (NA) where $k_y$/$k_0$ and $k_x$/$k_0$ are relative to normalized in-plane wavevectors, with $k_0$ the wavevector in free space. The maximum detectable in-plane wavevector is given by the NA of the objective and sets the outer rim radius of the Fourier plane at 1.49. The inner ring is located at NA = 1.0 and corresponds to the critical angle at the glass/air interface. The emission diagram of the N-PL feature an inhomogeneous distribution of the intensity indicating that some in-plane wavevectors bear more weight than others. This is particularly the case for the $+k_y$/$k_0$ wavevectors located near the detection limit. 

We quantitatively analyze this Fourier plane pattern by plotting radial cross-sections along the $k_y$/$k_0$ (at $k_x$/$k_0$ = 0) and $k_x$/$k_0$ axes (at $k_y$/$k_0$ = 0) shown as black and red curves in Fig.~\ref{fig:2}(b), respectively. The cross-sections indicate that intensity maximum is aligned along the $k_y$/$k_0$ axis with a peak at $k_y$/$k_0$ $\sim$ +1.43. As a reminder, we recall that the main axis of the AuNW is oriented along the $y$-axis. The black curve shows that the intensity in the $-k_y$/$k_0$ direction is smoothly distributed for $k_y$/$k_0>1.0$ with a small increase towards higher wavevectors in comparison to the $+k_y$/$k_0$ direction where the maximum intensity clearly peaks. We hypothesize that the asymmetric distribution of intensity along the $k_y$/$k_0$ axis is an effect of the asymmetry of the N-PL intensity distributed all along the antenna (Fig. \ref{fig:1}(c)) as discussed later on. The red curve shows that the intensity maxima along $k_x$/$k_0$ are located near the critical angle ($\pm k_x/k_0\sim$1.1). This feature is understood from the local excitation of the N-PL by the tightly focused excitation spot. We have quantified the directionality of the antenna $D = 10\log{I_{\rm{F}}/I_{\rm{B}}}$ along $k_y$/$k_0$ axis by measuring the intensity ratio between the point with maximum intensity $I_{\rm{F}}$ in the upper half and the diametrically opposite point with intensity $I_{\rm{B}}$ in the lower half of the Fourier plane. The $D$ value is 3.15 dB for intensity maximum at $k_y$/$k_0$ = +1.43.          

The Fourier plane highlights that the emitted rays are not only emerging at specific  $k_y$/$k_0$ wavevectors, but also features a limited azimuthal spread. We quantitatively evaluate the azimuthal extension by plotting the intensity as a function of the azimuthal angle, $\phi$, at NA = 1.1, 1.3 and 1.43, shown in Fig.~\ref{fig:2}(c). The azimuthal plots show that the maximum N-PL is angularly restricted and symmetric around the $k_y$/$k_0$ axis corresponding to the long-axis of the AuNW antenna. The intensity maxima are at $\phi$ $\sim$ 90$\degree$ ($-k_y$/$k_0$ axis) and $\phi$ $\sim$ 270$\degree$ ($+k_y$/$k_0$ axis) with the angular spread decreasing as NA increases in the lower half of the Fourier plane image.

In what follows, we provide an understanding of the different parts of the AuNW antenna contributing to the complex Fourier plane pattern quantified in Fig. \ref{fig:2}.

\begin{figure}[b]
\includegraphics{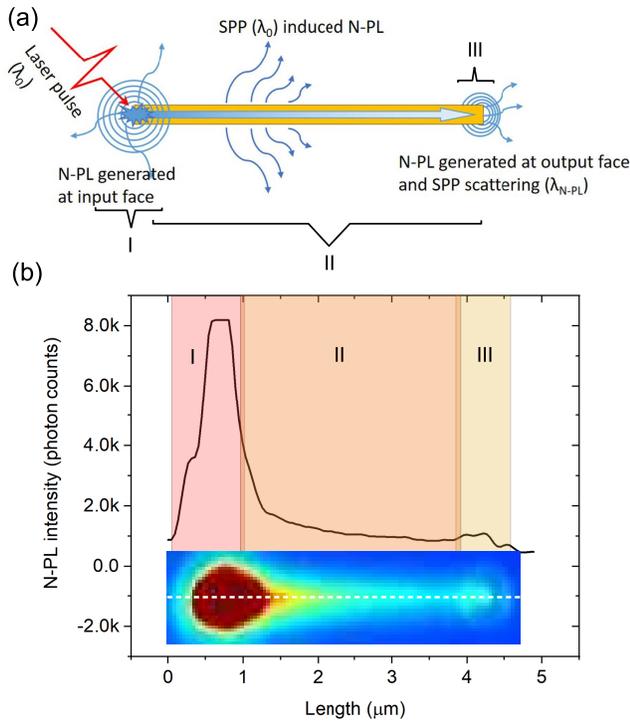}
\caption{\label{fig:3}(a) Schematic representation of the N-PL distributed in a AuNW. (b) Intensity cross-section along the AuNW indicates three different regions (I, II, and III) of the distributed N-PL in the AuNW. N-PL is generated at the input end of the AuNW by direct laser beam excitation (region I). SPP traveling in the AuNW at the excitation wavelength ($\lambda_0$ = 808nm) generates N-PL along the nanowire (region II). N-PL in region III is a sum of a local N-PL emission produced by the pump SPP scattering at the extremity and a continuum of SPP modes launched in region I within the N-PL spectrum, travelling and out-coupled by the distal end.}
\end{figure}

\subsection{Position dependent scattering of the N-PL local source by the AuNW antenna}

\subsubsection{Definition of the N-PL emissive regions}
The distribution of N-PL intensity across the AuNW is a result of multiple processes generating the nonlinear response~\cite{ref23}. Figure \ref{fig:3}(a) schematically pictures how the N-PL develops in a AuNW. Overlapping the input laser beam with a AuNW extremity results in large electric field at the dielectric discontinuity, leading to enhanced localized N-PL emission at the excitation point. This region is indicated as region I in Fig.~\ref{fig:3}(a). The excitation of the extremity launches a SPP propagating in the AuNW at the excitation wavelength (808 nm) as already illustrated in Fig.~\ref{fig:1}. If the intensity of the pump is large enough, the SPP triggers a nonlinear interaction as it propagates and results in the distributed N-PL along the AuNW, represented as region II. This delocalized N-PL signature is not a mode, but is locally produced at the metal surface by the underlying SPP traveling at $\lambda_0$~\cite{ref21,ref23}. When the SPP reaches the distal end, it produces a localized N-PL response indicated as region III. Region III also contains scatterring of a surface plasmon continuum. These secondary plasmons are launched at the input extremity by the strong and localized N-PL produced by the laser focus. They are excited with a continuum of wavelength contained within the N-PL emission spectrum (Fig.~\ref{fig:1}). These N-PL SPPs out-couple at $+k_y$/$k_0$ directions at higher wavevectors with a wide azimuthal spread~\cite{ref15}. This contribution is however not the predominant source of the N-PL photons at the distal end, since this set of secondary plasmons suffers from large propagation losses compared to the plasmon excited at the near infrared pump wavelength. Considering the N-PL produced at both extremities as two local secondary sources of light, the nanowire acts as antenna and scatters the source located in region III in the exact opposite way it scatters N-PL from region I because of the mirror symmetry. However, the intensity in region III remains much weaker than the other two regions (see Fig. \ref{fig:3}(b)), and we do not observe a clear symmetric $-k_y$/$k_0$ signature at -1.43.

Hence, understanding the contributions from the two remaining regions i.e., region I and II can provide information about the effect of different parts of the AuNW on the N-PL wavevector distribution presented in Fig. \ref{fig:2}(a). This analysis is only possible because N-PL is an incoherent process and the Fourier plane is not an interferogram.

\subsubsection{Fourier contributions from region I}

As illustrated in Fig. \ref{fig:3}, N-PL emission in region I regards the AuNW in the $+y$-axis and free space in the $-y$-axis. N-PL emission in the region II perceives the presence of AuNW in both directions. To separately study the wavevector distribution stemming from these two situations which, considering the relative weight of the nonlinear response at these positions, maximally contribute to the Fourier plane distribution in Fig. \ref{fig:2}(a), we exciting the AuNW at two different locations and for two incident polarization orientations.

\begin{figure}
\includegraphics{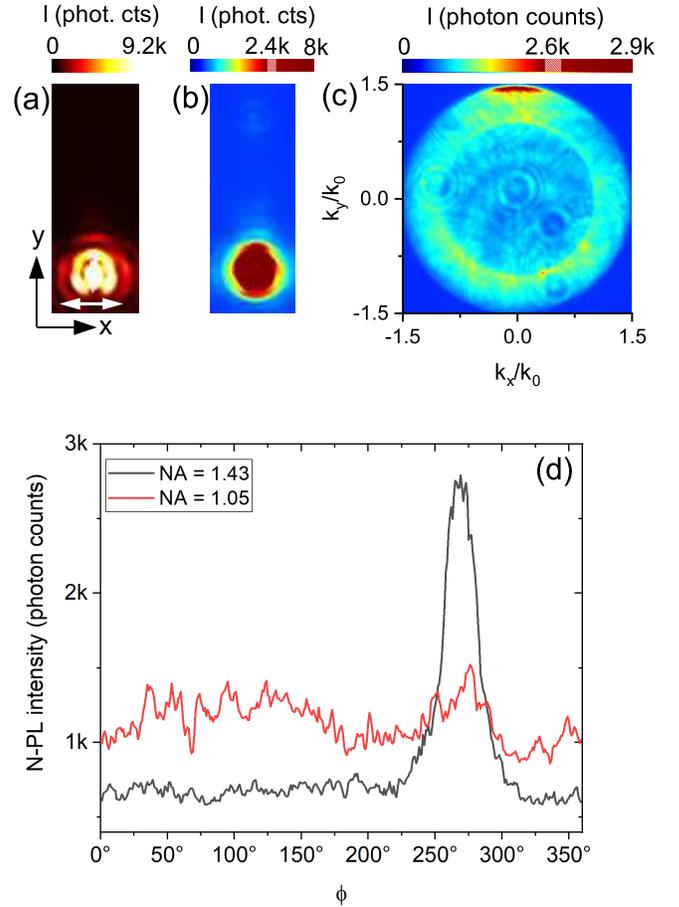}
\caption{\label{fig:4} (a) Wide-field real plane image taken at $\lambda_0$ for an excitation polarized transverse to the nanowire. No SPP is generated for this polarization and position of the nanowire in the focus. (b) and (c) Corresponding filtered N-PL wide-field real and Fourier plane images, respectively. (d) N-PL intensity distribution profile along the azimuthal coordinate ($\phi$) plotted at numerical aperture values NA = 1.05 and NA = 1.43. Small diffraction rings observed in the Fourier plane are from hard-to-remove stains at the surface of relay lenses.}
\end{figure}

Firstly, to extract the wavevector content of the local N-PL produced by the laser in the region I and scattered by the physical discontinuity, we excite one end of the AuNW with an input polarization transverse to its long-axis. In this configuration, coupling to a SPP at $\lambda_0$ remains inefficient~\cite{song} as illustrated in the wide-field real plane image of Fig. \ref{fig:4}(a) taken at the excitation wavelength. Hence, N-PL is generated essentially in region I and is absent from region II (Fig. \ref{fig:4}(b)). We observe an insignificant contribution in region III from N-PL-excited secondary SPP discussed above. 

The Fourier plane image in Fig.~\ref{fig:4}(c) shows the wavevector distribution of the N-PL for this excitation configuration. We observe a strong intensity at $+k_y$/$k_0$ near the detection limit, which strongly resembles the one measured in Fig.~\ref{fig:2}(a). This shared feature suggests that it is not the result of a SPP at $\lambda_0$. We claim that the reduction of the azimuthal and radial wavevector spread is the signature of an antenna effect brought by the quasi-one dimensional nanowire geometry. The directivity value calculated for the intensity maximum at $k_y$/$k_0$ = +1.43 is $D=5.6$~dB which is in the range of directionality achieved for optical Yagi-Uda antenna~\cite{ref12}.  

The Fourier plane in Fig.~\ref{fig:4}(c) contains also marked differences from Fig.~\ref{fig:2}(a). In particular, the intensity of the wavevectors located in the lower half of the diagram have a much larger azimuthal spread than the one measured for a polarization aligned with the nanowire. Azimuthal plots in Fig. \ref{fig:4}(d) indicate broader wavevector spread at NA = 1.05 and negligible intensity at NA = 1.43 in the lower half of the Fourier plane. This can be understood from the absence of the antenna in the $-y$ axis to the excitation spot and the absence of N-PL in region II.

\subsubsection{Fourier contributions from Region II}

\begin{figure}
\includegraphics{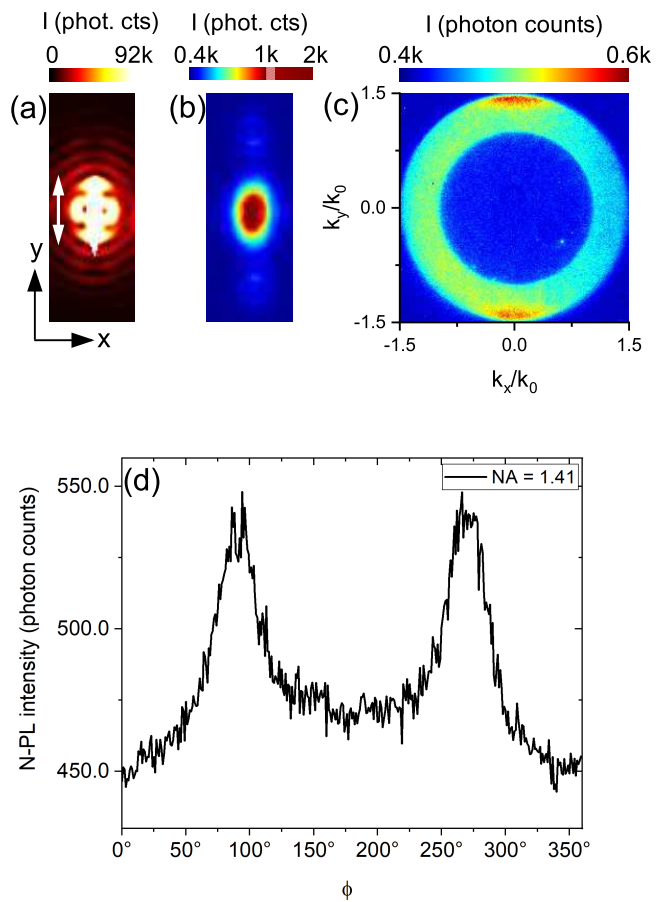}
\caption{\label{fig:5} (a) Wide-field real plane image taken at $\lambda_0$ for an excitation located at the center of the AuNW. Polarization is along the main axis (double arrow). SPP excitation is absent. (b) and (c) Corresponding filtered N-PL 
wide-field real plane and Fourier plane images, respectively. (d) N-PL intensity distribution profile along the azimuthal coordinate ($\phi$) plotted at numerical aperture value NA = 1.41.}
\end{figure}

Now, we try to understand the Fourier contribution to N-PL emitted from Region II of the nanowire. To remove the influence of the strong local response of the extremity, we move the laser beam to the center of the AuNW and align the incident polarization along the main axis. Here again, no SPP is launched in the AuNW because there is no momentum transfer provided by the geometry along the polarization direction. This is pictured in the wide-field image of Fig.~\ref{fig:5}(a) taken at the laser wavelength. In the laser-filtered image of Fig. \ref{fig:5}(b), the N-PL spatial extension is thus limited to the focus size with a negligible coupling of N-PL-excited secondary SPP to the nanowire. This situation mimics the effect of the nanowire on the N-PL generated locally along the AuNW surface at the center of region II (Fig. \ref{fig:3}).  As expected the presence of the nanowire antenna on both sides of the localized N-PL source renders a symmetric Fourier plane. This is pictured in Fig.~\ref{fig:5}(c), where the wavevector intensity peaks at the detection limit on both $+k_y$/$k_0$ and $-k_y$/$k_0$ axis. The angular azimuthal spreads are narrow on both $+k_y$/$k_0$ and $-k_y$/$k_0$ axis with an average FWHM = 40$^\circ$ as shown in the cross-sectional profile plotted at NA = 1.41 in Fig.~\ref{fig:5}(d).

We can summarize the explanation of the Fourier plane of distributed N-PL (Fig. \ref{fig:2}) using the above observations. The maximum intensity towards $+k_y$/$k_0$ is due to the dominant N-PL emitted in region I and redirected there by the scattering influence of the nanowire. A second contributor to this intensity maximum originates from the N-PL present in region II and, to a much lesser extent, the secondary SPPs launched by the local N-PL at the input and scattering at region III from the distal end.
Inplane wavevectors emerging at numerical aperture values near NA $\sim$ 1 in the lower half of Fourier plane are wavevectors which are not affected by nanowire antenna. N-PL filling the Fourier space above the critical angle in the $-k_y$/$k_0$ direction comes from region II complemented by the weak scattering of the localized N-PL source present at the distal end (region III). 

\subsection{Polarization analyzed wavevector distribution of the distributed N-PL}
We complete our measurements by analyzing the polarization of the emitted N-PL. We record the N-PL intensity distributions in real and Fourier spaces for two polarization orientations of the emitted N-PL. Polarization of the input beam remains fixed along the long-axis of the AuNW, and the analyzer axis is rotated either along or transverse to the long-axis of the AuNW. The excitation stays fixed at the lower extremity of the AuNW.

\begin{figure}
\includegraphics{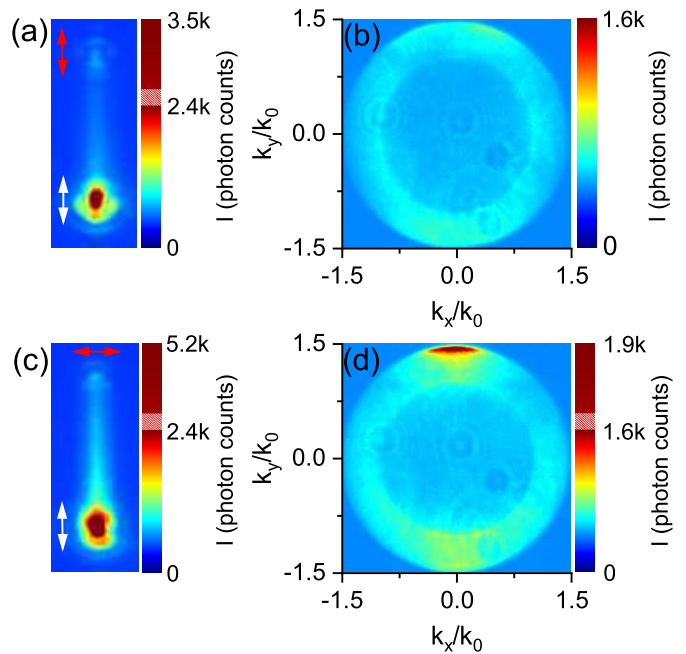}
\caption{\label{fig:6} Polarization analyzed ((a) and (c)) wide field real plane images and ((b) and (d)) Fourier plane images. White and red arrows indicate the axis of polarization of the input laser beam and of the analyzer, respectively.}
\end{figure}

Figure \ref{fig:6}(a) and (b) show a wide-field real plane image and the corresponding Fourier plane image of the N-PL polarized along the long-axis of the AuNW, respectively. While the wide-field image does not qualitatively differs from Fig.~\ref{fig:1}, the Fourier plane is now essentially symmetric with respect to the $k_y$/$k_0$ and $k_x$/$k_0$ axes. The narrowing of the wavevector spread in the radial and azimuthal directions at $+k_y$/$k_0\sim1.43$ imparted by the nanowire is absent. We understand this from the following argumentation. We assume for the sake of the argument that the local sources positioned at either extremity and along the nanowire can be considered as a random distribution of oriented dipoles. The orientation of the analyzer selects the dipole moments preferentially emitting along the nanowire. In far-field, these dipole emission diagrams features the well-known two-lobe pattern oriented perpendicularly to the dipole moment (and the axis of the nanowire). This particular orientation of the emission lobes mitigate therefore the antenna effect.

Now, the analyzer axis is rotated transverse to the long-axis of the AuNW to analyze the N-PL polarized along this direction. The results are displayed in the wide-field real and Fourier plane images in Fig. \ref{fig:6}(c) and (d). Again, the spatial distribution of the N-PL closely resembles those already observed for the cross-polarization detection and for the unpolarized detection (Fig.~\ref{fig:1}). Quantitatively, the intensity of the N-PL in region II is higher than in the previous polarization detection, mainly because the efficiency of the nonlinear process is helped by an interaction with the surface of the metal~\cite{ref23}.

The Fourier plane essentially features all the important details already observed in Fig.~\ref{fig:1}. Considering again the local N-PL sources at the extremities and along the nanowire as dipoles oscillating perpendicularly to the nanowire axis, this particular orientation of the analyzer favors an emission diagram aligned with the AuNW, maximizing henceforth the antenna effect, and concentrating the wavevector content to a narrow angular distribution at high $+k_y$/$k_0$.

\subsection{Simulation results}

\begin{figure}[b]
\includegraphics{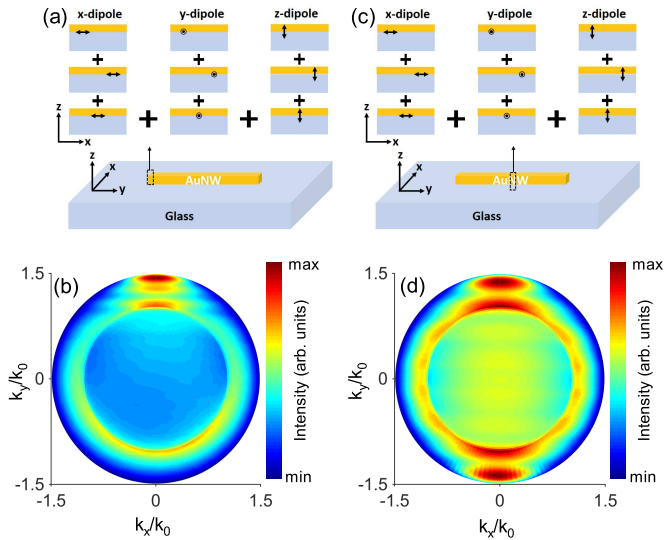}
\caption{\label{fig:7} (a) and (c) Schematic placement of dipoles at the end and at the center of the AuNW, respectively. Fourier plane images are simulated for single dipole placed at positions $x$ = 0, $-40$~nm and $+40$~nm across AuNW width for wavelength range 550 nm – 680 nm in 10 nm steps. (b) and (d) Fourier plane images representing the incoherent sum for the dipoles placed at the extremity (a) and the center of the nanowire (c), respectively.}
\end{figure}

We have performed a series of simulations to confirm the experimental N-PL wavevector distributions. Three-dimensional finite element method is used to calculate the near-field electric field using Comsol Multiphysics software. Far-field wavevector intensity distribution of the fields emitted into the substrate are calculated by an analytical treatment based on reciprocity arguments~\cite{ref26}. The AuNW is modeled as a 3.5 $\mu$m long, 160 nm wide and 50 nm tall cuboidal structure. The cuboid is placed on a glass substrate (refractive index = 1.50) and is covered with air (refractive index = 1). The Au wavelength dependent refractive index is taken from~\cite{ref25}. Top edges of the cuboid are rounded with a 5 nm radius to reproduce the definition of lithographically produced AuNW. A dipole source is positioned at one end of the nanowire as schematically shown in Fig. \ref{fig:7}(a). The extent of N-PL along the width is mimicked by positioning dipoles at positions $x$ = 0, $-40$~nm and $+40$~nm along the width of the AuNW at the glass/metal interface. Fourier plane patterns are simulated for dipoles with equal strength, oscillating along $x$ (transverse to the long axis of the nanowire), $y$ (along the long axis of the nanowire) and $z$ (out of plane) axes in the N-PL wavelength range (550 nm – 680 nm).  

The N-PL is an incoherent process and its wavevector distribution emitted in the substrate may be modeled as an incoherent sum of Fourier plane patterns obtained for dipoles oscillating in $x, y$ and $z$ axes within the N-PL wavelength range:
\begin{eqnarray}
I_{\rm{N-PL}}(k_x,k_y) = \sum_{\lambda}a_\lambda [I_{x}(k_x,k_y,\lambda)\nonumber\\
+ I_{y}(k_x,k_y,\lambda) + I_{z}(k_x,k_y,\lambda)]
\label{equ:1}
\end{eqnarray}

Here $a_\lambda$ is a coefficient that includes the weight of the N-PL at different wavelengths taken from the N-PL spectrum (Fig.~\ref{fig:1}) and $I_{(x,y,z)}$ are the calculated intensity diagrams for each dipole axes. The resultant sum of $I_{\rm{N-PL}}(k_x,k_y)$ for different positions along $x$ axis in Fig. \ref{fig:7}(b) resembles the experimentally observed wavevector distribution in Fig. \ref{fig:4}(c). The salient features observed experimentally are reproduced in the simulations, especially the distribution of intensity peaking along the $+k_y$/$k_0$ axis. Additional fringes not observed experimentally come from the coherent nature of dipole sources at a particular wavelength and for a particular polarization in the simulations. In contrast, N-PL by the nature of its generation is an incoherent emission process~\cite{remesh}.

Next the probing dipole is moved to the center (lengthwise) of the AuNW, as shown in Fig. \ref{fig:7}(c), and the same simulations as above are repeated to mimic the Fourier plane pattern of the N-PL in Fig. \ref{fig:5}(c). In this configuration dipoles perceive the presence of nanowire in both $-y$ and $+y$ directions and hence have a symmetric distribution along the $k_y$/$k_0$ axis (Fig. \ref{fig:7}(d)). The resultant incoherent sum of Fourier plane patterns in Fig. \ref{fig:7}(d) is in qualitative agreement with the experimental distribution directions in Fig. \ref{fig:5}(c). Hence the simulation results qualitatively support the experimental observation that different parts of the nanowire antenna generate N-PL in-plane wavevectors differently and confirms our hypothesis.\\ 

\begin{figure}
\includegraphics{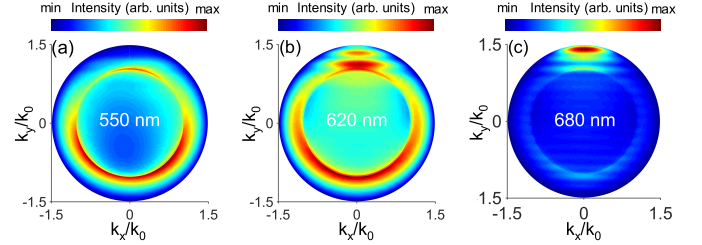}
\caption{\label{fig:8} (a) - (c) Calculated Fourier plane images for the incoherent dipole source placed at the end of the AuNW at single wavelengths 550 nm 620 nm and 680 nm respectively.
}
\end{figure}

The wavevector distribution of N-PL depends on the detection window of the spectrum \cite{ref21}. For qualitative understanding of wavelength dependent wavevector distribution of N-PL, we have calculated Fourier plane images for incoherent dipole source at the end of the AuNW (Fig \ref{fig:7}(a)) at single wavelengths 550 nm, 620 nm and 680 nm in Fig. \ref{fig:8}(a-c). Observed wavevector distribution differs from each other and Fourier plane images suggest that intensity maximum observed in upper half of experimental Fourier plane image has larger weight of higher wavelengths. This feature can be understood from material dispersion which prevents the antenna to operate efficiently for shorter wavelengths.

\section{Conclusion}
Using a quantitative analysis of Fourier plane patterns, we have demonstrated the effect of the AuNW antenna on the directionality of N-PL. Distributed N-PL over the AuNW results in an intricate Fourier plane pattern, which is explained by exciting different parts of antenna separately. Our polarization analyzed study suggests that the directional signal of the distributed N-PL, mapped at higher wavevectors, is maximally polarized transversely to the long-axis of the antenna. Finite element method based simulations confirm that the directionality of the radiation pattern depends on the placement of the incoherent source on different positions on the nanowire antenna. Our simulations results have considered N-PL source mimicked by dipoles, and are in qualitative agreement with our experimental results. Hence, our study of wavevector analysis may be equally applied to N-PL generated by more complex structures than a nanowire.

\begin{acknowledgments}
The work was made possible through the Indo-French Centre for the Promotion of Advanced Research, project No. 5504-3, the project APEX funded by the Conseil R\'egional de Bourgogne Franche-Comt\'e, the European Regional Development Fund (ERDF), EUR-EIPHI (17-EURE-0002). Device fabrication was performed in the technological platform ARCEN Carnot with the support of the R\'egion de Bourgogne Franche-Comt\'e and the D\'el\'egation R\'egionale \`a la Recherche et \`a la Technologie (DRRT). Calculations were partially performed using HPC resources from DSI-CCuB (Université de Bourgogne). GVPK acknowledges DST, India for Swarnajayanti fellowship grant (DST/SJF/PSA-02/2017–18). DKS thanks Adarsh B. Vasista and Diptabrata Paul for fruitful discussions.
\end{acknowledgments}

\nocite{*}

%

\end{document}